\documentclass[aps,prd,twocolumn,superscriptaddress]{revtex4-2}
\newcommand{\jcap}{J. Cosmol. Astropart. Phys. }

\newcommand{\apjl}{Astrophys. J. Lett.}
\newcommand{\physrep}{Phys. Rep. }
\newcommand{\aap}{Astron. Astrophys. }

\newcommand{\mnras}{Mon. Not. R. Astron. Soc. }

\newcommand{\pasj}{Publ. Astron. Soc. Japan }
\newcommand{\nphysa}{Nucl. Phys. A }
\usepackage{color}

\usepackage{graphicx}
\newcommand{\eda}{\mathcal{E}}
\newcommand{\fla}{\mathcal{F}}

\begin{document}

\title{Shock Revival in Core-collapse Supernovae Assisted by Heavy Axion-like Particles}

\author{Kanji Mori}
\email[]{kanji.mori@fukuoka-u.ac.jp}
\affiliation{Research Institute of Stellar Explosive Phenomena, Fukuoka University, 8-19-1 Nanakuma, Jonan-ku, Fukuoka-shi, Fukuoka 814-0180, Japan}
\author{Tomoya Takiwaki}
\affiliation{National Astronomical Observatory of Japan, 2-21-1 Osawa, Mitaka, Tokyo 181-8588, Japan}
\author{Kei Kotake}
\affiliation{Research Institute of Stellar Explosive Phenomena, Fukuoka University, 8-19-1 Nanakuma, Jonan-ku, Fukuoka-shi, Fukuoka 814-0180, Japan}
\affiliation{Department of Applied Physics, Faculty of Science, Fukuoka University, 8-19-1 Nanakuma, Jonan-ku, Fukuoka-shi, Fukuoka 814-0180, Japan}
\author{Shunsaku Horiuchi}
\affiliation{Center for Neutrino Physics, Department of Physics, Virginia Tech, Blacksburg, VA 24061, USA}
\affiliation{Kavli IPMU (WPI), UTIAS, The University of Tokyo, Kashiwa, Chiba 277-8583, Japan}

\date{\today}

\begin{abstract}
Axion-like particles (ALPs) are a class of hypothetical pseudoscalar particles which feebly interact with ordinary matter. The hot plasma of core-collapse supernovae is a possible laboratory to explore physics beyond the standard model including ALPs. Once produced, some of the ALPs can be absorbed by the supernova matter and affect energy transfer. In this study, we calculate the ALP emission in core-collapse supernovae and the backreaction on supernova dynamics consistently. It is found that the stalled bounce shock can be revived if the coupling between ALPs and photons is as high as $g_{a\gamma}\sim 10^{-9}$ GeV$^{-1}$ and the ALP mass is 40-400 MeV. {Most of the models result in more energetic explosions than the average observed supernova. While this can be used to place constraints on those ALPs, long-term simulations across multiple progenitors need to be further investigated to place robust limits.}
\end{abstract}

\maketitle

\section{Introduction}
Axion-like particles (ALPs) are exotic pseudoscalar particles that possibly interact with photons \citep[e.g.][]{2020PhR...870....1D,2020arXiv201205029C}. The ALP-photon interaction is described by the Lagrangian \citep{1988PhRvD..37.1237R}
\begin{equation}
    \mathcal{L}=-\frac{1}{4}g_{a\gamma}F_{\mu\nu}\tilde{F}^{\mu\nu}a,\label{lag}
\end{equation}
where $a$ is the ALP field, $F_{\mu\nu}$ is the electromagnetic tensor and $\tilde{F}_{\mu\nu}$ is its dual, and $g_{a\gamma}$ is the coupling constant between ALPs and photons. The coupling with standard model particles leads to the production of ALPs in astrophysical plasma.

Heavy ALPs with a mass of $m_a\gtrsim 1$\,MeV have been explored by experiments and cosmological and astrophysical arguments. Terrestrial experiments with particle accelerators have excluded a large part of the ALP parameter space with $g_{a\gamma}\gtrsim10^{-7}$\,GeV$^{-1}$ \citep{2016PhLB..753..482J,2017JHEP...12..094D,2019JHEP...05..213D,2020PhRvL.125h1801B}. Also, standard cosmology including Big Bang nucleosynthesis and the cosmic microwave background provides strong constraints on the ALP parameters \citep{2012JCAP...02..032C,2020JCAP...05..009D}.

Astrophysical arguments often utilize stars and supernovae (SNe) as ALP factories. Since ALPs produced in stars affect the energy transfer, the parameters can be constrained by comparison between stellar models and astronomical observations. For example, additional energy losses from stars induced by ALPs shorten the lifetime of horizontal branch stars in globular clusters and affect the stellar population \citep{2020PhLB..80935709C}. The energy loss also changes the structure of asymptotic giant branch stars and leads to a different initial-final mass relation of white dwarfs \citep{2021arXiv210200379D}. Many ALPs can also be produced in core-collapse SNe (CCSNe). As a result, stringent constraints can be obtained from the neutrino burst from SN 1987A \citep{2018arXiv180810136L,2020JCAP...12..008L} and the explosion energy \cite{2019PhRvD..99l1305S,2022arXiv220109890C}. A part of heavy ALPs can escape from astronomical objects and decay into photons during propagation in the interstellar space. Non-detection of $\gamma$-rays from SN 1987A then gives another constraint on the ALP mass and $g_{a\gamma}$. These astrophysical considerations constrain the ALP-photon coupling to approximately $g_{a\gamma}\lesssim10^{-10}$--$10^{-9}$\,GeV$^{-1}$ across a broad range of ALP mass. In the future, $\gamma$-rays from nearby core-collapse and thermonuclear SNe and hypernovae \citep{2011JCAP...01..015G,2018PhRvD..98e5032J,2021PASJ..tmp...91M,2021arXiv210405727C} may lead to additional constraints.  
 
In previous works of CCSNe, the ALP production and the hydrodynamics were decoupled from each other. Although one can post-process the calculation of the the ALP luminosity, the backreaction on SN dynamics cannot be investigated in this way. Nevertheless, {Refs.~\citep{2020JCAP...12..008L,2021arXiv210903244C}} recently pointed out that heating in the SN gain region due to heavy ALPs can reach $\sim10^{52}$\,erg s$^{-1}$ if the ALP mass is $m_a\sim200$\,MeV. In one-dimensional SN models, the shock stalls and a successful explosion cannot be obtained (see e.g., Refs.~\citep{2018JPhG...45j4001O,2021Natur.589...29B} for  reviews). However, it has been reported that beyond Standard Model particles such as radiatively decaying ALPs \cite{1982ApJ...260..868S} and sterile neutrinos \cite{2018PhRvD..98j3010R} may cause additional heating of the shock that can lead to shock revival and significantly affect the explosion mechanism. Also, since a part of the neutrino energy  may be carried out by ALPs, neutrino and gravitational wave signals from an SN may be altered. It is therefore desirable to investigate the backreaction on hydrodynamics  self-consistently.
 
In this {study}, we perform a series of one-dimensional hydrodynamical simulations of stellar core collapse including both ALP production and their backreactions. We show that depeding on the ALP mass and its photon coupling, we can obtain shock revival due to axion energy transport in a progenitor which does not explode when ALPs are not included. 

{This paper is organized as follows. In Section II and III, the prescription for the ALP production and absorption rates is discussed, respectively. In Section IV, we explain the setup for our SN models and the implementation of ALP heating. In Section V, we show the results of our calculations. In Section VI, we summarize our results and discuss future prospects.}
 
 \section{ALP production rates}
 
 In this study, we consider photophilic ALPs that interact with photons.  We take into account two processes for ALP production: the Primakoff process $(\gamma+p\rightarrow a+p)$ catalysed by protons and the photon coalescence  $(\gamma+\gamma\rightarrow a)$. 
 
 The Primakoff rate is given as \citep{2000PhRvD..62l5011D}
 \begin{eqnarray}
     \Gamma_{\gamma\rightarrow a}=g_{a\gamma}^2\frac{T\kappa^2}{32\pi}\frac{p}{E}
     \left(\frac{((k+p)^2+\kappa^2)((k-p)^2+\kappa^2)}{4kp\kappa^2}\times\right.\nonumber\\
     \left.\ln\left(\frac{(k+p)^2+\kappa^2}{(k-p)^2+\kappa^2}\right)
-\frac{(k^2-p^2)^2}{4kp\kappa^2}\ln\left(\frac{(k+p)^2}{(k-p)^2}\right)-1\right),\label{prim}
 \end{eqnarray}
 where $T$ is the temperature, $E$ is the ALP energy, $p=\sqrt{E^2-m_a^2}$ is the ALP momentum,  $k=\sqrt{\omega^2-\omega_\mathrm{pl}^2}$ is the  wave number of photons in plasma, $\omega$ is the photon energy, $\omega_\mathrm{pl}\approx16.3$ MeV $Y_e^\frac{1}{3}(\rho/10^{14}\;\mathrm{g\;cm}^{-3})^\frac{1}{3}$ is the plasma frequency \citep{1998PhRvD..57.3235K}, $Y_e$ is the electron mole fraction, $\kappa=\sqrt{4\pi\alpha n_p^\mathrm{eff}/T}$ is the Debye-H\"{u}ckel scale, $\alpha\approx1/137$ is the fine structure constant, and $n_p^\mathrm{eff}$ is the effective proton number density. The energy conservation leads to $E=\omega$. The  proton number density with the Pauli blocking effect is defined as
 \begin{eqnarray}
 n_p^\mathrm{eff}=2\int\frac{d^3\mathbf{p}}{(2\pi)^3}f_p(1-f_p),
 \end{eqnarray}
 where $f_p$ is the Fermi-Dirac distribution of protons that depends on the proton chemical potential and the effective proton mass in plasma. The chemical potential is determined from the relation
 \begin{eqnarray}
 n_p=2\int\frac{d^3\mathbf{p}}{(2\pi)^3}f_p,
 \end{eqnarray}
 where $n_p$ is the number density of protons. The effective proton mass is calculated as \citep{1999PhRvC..59.2888R,2006A&A...447.1049B}
 \begin{eqnarray}
 m_p^\ast(\rho)=\frac{m_p}{1+\frac{a\rho}{\rho_\mathrm{nuc}}},
 \end{eqnarray}
 where $m_p\approx938$ MeV is the proton mass, $\rho_\mathrm{nuc}=0.16m_p$ fm$^{-3}$ is the saturation density, and $a$ is a constant that is determined by an equation $m_p^\ast(\rho_\mathrm{nuc})=0.8m_p$.

 The energy loss rate induced by the Primakoff process is written as
 \begin{eqnarray}
 Q_\mathrm{cool}&=&\int^\infty_{m_a}dEE\frac{d^2n_a}{dtdE}
=2\int\frac{d^3\mathbf{k}}{(2\pi)^3}\Gamma_{\gamma\rightarrow a}\omega f(\omega),\label{q1}
 \end{eqnarray}
 where $f(\omega)$ is the Bose-Einstein distribution of photons.
 
 When ALPs are heavier than $2\omega_\mathrm{pl}$, the photon coalescence contributes to the ALP production. The  photon coalescence rate is given as \cite{2000PhRvD..62l5011D}
 \begin{eqnarray}
 \frac{d^2n_a}{dtdE}=g_{a\gamma}^2\frac{m_a^4}{128\pi^3}p\left(1-\frac{4\omega_\mathrm{pl}^2}{m_a^2}\right)^\frac{3}{2}e^{-\frac{E}{T}}.\label{pc}
 \end{eqnarray}
 The energy loss rate due to the photon coalescence is then
 \begin{eqnarray}
 Q_\mathrm{cool}=\int^\infty_{m_a}dEE\frac{d^2n_a}{dtdE}.\label{q2}
 \end{eqnarray}

\section{ALP Absorption Rates}

Once produced, ALPs propagate through the SN matter. If they are absorbed by the matter during propagation, they affect the energy transfer in the SN. In this study, we take into account the inverse Primakoff process $(a\rightarrow \gamma)$ and the radiative decay $(a\rightarrow\gamma\gamma)$ as ALP absorption processes \cite{2020JCAP...12..008L}.
 
The inverse Primakoff rate is written as $\Gamma_{a\rightarrow \gamma}=2\Gamma_{\gamma\rightarrow a}/\beta_E$. The mean free path (MFP) of ALPs due to the inverse Primakoff process is given by  $\lambda_{a\rightarrow \gamma}=\beta_E\gamma_E/\Gamma_{a\rightarrow \gamma}$, where $\gamma_E$ is the Lorentz factor of ALPs and $\beta_E=\sqrt{1-\gamma_E^{-2}}$. The radiative decay rate is estimated as
 \begin{eqnarray}
 \Gamma_{a\rightarrow \gamma\gamma}=g_{a\gamma}^2\frac{m_a^3}{64\pi}\left(1-\frac{4\omega_\mathrm{pl}^2}{m_a^2}\right)^\frac{3}{2}. \label{decay}
 \end{eqnarray}
The MFP from this process is given as $\lambda_{a\rightarrow \gamma\gamma}=\beta_E\gamma_E/\Gamma_{a\rightarrow \gamma\gamma}$.
The total MFP is then given by $\lambda_a=(\lambda_{a\rightarrow \gamma}^{-1}+\lambda_{a\rightarrow \gamma\gamma}^{-1})^{-1}$. Although the MFP is dependent on ALP energy $E$, we average $E$ over the ALP spectrum to reduce the computational cost.

 \section{Core collapse models}

 We incorporate the effect of ALPs in \texttt{3DnSNe} \cite{2016MNRAS.461L.112T} and perform one-dimensional SN hydrodynamical simulations. The code adopts HLLC solver \cite{toro94} to solve the Riemann problem. The nuclear equation of state is based on Ref.~\cite{1991NuPhA.535..331L} with incompressibility of 220\,MeV. The neutrino transport is treated with a three-flavor isotropic diffusion source approximation \cite{2009ApJ...698.1174L,2014ApJ...786...83T,2018ApJ...853..170K}. The employed neutrino reactions are the same as set6abc of Ref.~\cite{2018ApJ...853..170K}.
 We adopt {non-rotating progenitors \cite{2002RvMP...74.1015W} with the zero-age main sequence (ZAMS) masses of $20M_\odot$ and $11.2M_\odot$ and the Solar metallicity as the initial conditions. Although a thorough sensitivity study on the progenitor mass dependence is beyond the scope of this work, the two progenitors are adopted because they are typical models that would leave a neutron star as an outcome of the core-collapse \cite[e.g.][]{2016ApJ...821...38S}}.

 At the final stage of the evolution of massive stars, photodisintegration of iron destabilizes the stellar core and  core collapse commences. When the density of the collapsed core reaches the nuclear saturation density, the core stiffens due to the nuclear repulsive force and the core bounce occurs.  A shock is then formed in the core because the SN material continues to accrete from the envelope.  The temperature is highest at a radius of $\sim10$ km because the shock there heats the accreting material. 
 Eventually, the shock wave that is formed by the core bounce stalls because of energy losses due to the photodisintegration of heavy elements.
 In our model without ALPs, the stalled shock is not energetically revived, consistent with the literature where one-dimensional explosions are achieved only in lighter $\sim 8$--$10 M_\odot$ stars \cite{2006A&A...450..345K}.

 In order to incorporate the effects of ALPs in SN models, we treat the ALP transport as follows. The evolution of the ALP energy per unit volume $\eda$ is described by 
\begin{eqnarray}
\frac{\partial \eda }{\partial t}
+\nabla\cdot \mathbf{\fla}
=Q_\mathrm{cool}-Q_\mathrm{heat},
\end{eqnarray}
 where $\mathbf{\fla}$ is the ALP energy flux and $Q_\mathrm{heat}$ is the heating rate per unit volume due to ALPs. If we assume stationarity and spherical symmetry, the equation simplifies to
 \begin{eqnarray}
 \frac{1}{4\pi r^2}\frac{\partial}{\partial r}(L)=Q_\mathrm{cool}-Q_\mathrm{heat},\label{eq3.2}
 \end{eqnarray}
 where $L=4\pi r^2\fla$.
Eq.~(\ref{eq3.2}) is discretized in radius as
 \begin{eqnarray}
 L_{i+\frac{1}{2}}=L_{i-\frac{1}{2}}+(Q_{\mathrm{cool},\;i}-Q_{\mathrm{heat},\;i})\Delta V_i,\label{rec}
 \end{eqnarray}
 for the $i$-th cell. Here $ L_{i+\frac{1}{2}}$ and $ L_{i-\frac{1}{2}}$ are the ALP luminosities at the cell edges and $\Delta V_i$ is the cell volume. The heating rate $Q_\mathrm{heat}$ is evaluated as
 \begin{eqnarray}
 Q_{\mathrm{heat},\;i}\Delta V_i=L_{i-\frac{1}{2}}\left(1-\exp\left(-\frac{r_{i+1}-r_{i}}{\lambda_{a,\;i}}\right)\right).\label{qheat}
 \end{eqnarray}
 The $n$-th step of the internal energy of the matter, $e_\mathrm{int}^{n}$,  is then updated as
\begin{eqnarray}
e_{\mathrm{int},\;i}^{n+1} = e_{\mathrm{int},\;i}^n +(Q_{\mathrm{heat},\;i}^n -Q_{\mathrm{cool},\;i}^n)\Delta t,
\end{eqnarray}
where $\Delta t$ is the time step. 

Similarly, the number $L_{\mathrm{n}}$ of ALPs that pass a mass shell per a unit time follows the relation
\begin{eqnarray}
 L_{\mathrm{n},\;i+\frac{1}{2}}=L_{\mathrm{n},\;i-\frac{1}{2}}+(\dot{N}_{\mathrm{cool},\;i}-\dot{N}_{\mathrm{heat},\;i})\Delta V_i,
\end{eqnarray}
where $\dot{N}_{\mathrm{cool},\;i}$ $(\dot{N}_{\mathrm{heat},\;i})$ is the number of ALPs produced (absorbed) in the $i$-th cell per a unit time. The number of absorbed ALPs is estimated by
\begin{eqnarray}
 \dot{N}_{\mathrm{heat},\;i}\Delta V_i=L_{\mathrm{n},\;i-\frac{1}{2}}\left(1-\exp\left(-\frac{r_{i+1}-r_{i}}{\lambda_{a,\;i}}\right)\right).\label{nheat}
\end{eqnarray}
The average ALP energy $E_{\mathrm{ave},\;i}$ in the $i$-th cell is then estimated as $E_{\mathrm{ave},\;i}=L_{i-\frac{1}{2}}/L_{\mathrm{n},\;i-\frac{1}{2}}$.
The averaged energy is used to calculate the MFP of ALPs.

\section{Results}
\begin{table*}[]
\begin{tabular}{ccc|cccccc}
$M/M_\odot$&$m_a$ [MeV] & $g_{10}$ & Shock revival? &$t_\mathrm{pb,\;400}$ [ms]& $E_\mathrm{exp}$ {[}$10^{51}$ erg{]} & $M_\mathrm{PNS}/M_\odot$&$L_a$ [erg s$^{-1}$]&${L_\mathrm{heat}}$ [erg s$^{-1}$] \\\hline\hline
 20&  $-$   & 0        & No  &&           &                                     &                0        &0  \\\hline
20&50    & 4        & No  &&           &                                     &$3.94\times10^{50}$  &$3.88\times10^{47}$                        \\
20&50    & 10       & No    &&         &                                     &   $2.47\times10^{51}$   &$1.50\times10^{49}$                    \\
20&50    & 20       & No      &&       &                                     &   $9.90\times10^{51}$  &$2.38\times10^{50}$                     \\
20&50    & 40       & Yes    &557        &0.95                                     & 1.88  &$3.99\times10^{52}$ &$3.70\times10^{51}$                      \\\hline
20&100   & 4        & No  &&           &                                     &  $1.28\times10^{51}$    &$1.58\times10^{49}$                    \\
20&100   & 10       & No   &&          &                                     &  $7.98\times10^{51}$ &$6.00\times10^{50}$                       \\
20&100   & 20       & Yes      &352      &1.23                                     &1.85&$3.18\times10^{52}$  &$8.50\times10^{51}$                        \\
20&100   & 40       & Yes      &166      & 2.11& 1.69 &$1.05\times10^{53}$&$7.55\times10^{52}$                        \\\hline
20&200   & 4        & No      &&       &                                     & $3.28\times10^{51}$ &$4.75\times10^{50}$                        \\
20&200   & 10       & Yes      &308      &1.41                                     &   1.83  &$2.02\times10^{52}$ &$1.26\times10^{52}$                    \\
20&200   & 20       & Yes      &176      & 0.93                                    &  1.71   &$6.52\times10^{52}$& $6.37\times10^{52}$                    \\
20&200   & 40       & Yes        &134    & 0.99                                    & 1.63  &$1.74\times10^{53}$              &$1.73\times10^{53}$         \\\hline
20&400   & 4        & Yes     &526       & 0.40                                   & 1.90  &$1.32\times10^{51}$&$1.08\times10^{51}$                       \\
20&400   & 10       & Yes     &286       &  0.41                                   & 1.82 &$8.12\times10^{51}$ &$7.95\times10^{51}$                       \\
20&400   & 20       & Yes    &252        & 0.88                              &     1.78  &$3.10\times10^{52}$  &$3.03\times10^{52}$                      \\
20&400   & 40       & Yes      & 271     &1.11                                     &  1.79  &$1.09\times10^{53}$ &$1.05\times10^{53}$                     \\\hline
20&800   & 4        & No        &&     &                                     &  $8.48\times10^{48}$ &  $8.48\times10^{48}$                     \\
20&800   & 10       & No         &&    &                                     &  $5.30\times10^{49}$   &$5.30\times10^{49}$                     \\
20&800   & 20       & No        &&     &                                     &    $2.12\times10^{50}$ &     $2.12\times10^{50}$                \\
20&800   & 40       & No       &&      &                                     &  $8.48\times10^{50}$   &$8.48\times10^{50}$         \\\hline           
11.2&  $-$   & 0        & No     &&        &                                     &            0            &0  \\\hline
11.2&100   & 4        & No   &&          &                                     &   $3.38\times10^{50}$   &$4.70\times10^{48}$                    \\
11.2&100   & 10       & No        &&     &                                     & $2.10\times10^{51}$  &$1.76\times10^{50}$                       \\
11.2&100   & 20       & Yes  &358          &0.24                                     &1.35 & $8.08\times10^{51}$  &$2.39\times10^{51}$                     \\
11.2&100   & 40       & Yes &122           & 0.22                                    & 1.30& $2.30\times10^{52}$ &$1.77\times10^{52}$               

\end{tabular}
\caption{The SN models developed in this study. $M$ is the progenitor ZAMS masses  and $t_\mathrm{pb,\;400}$ is the time since the shock bounce at the moment when the shock  wave reaches $r=400$ km. The values of the explosion energy $E_\mathrm{exp}$ and the proto-neutron star mass $M_\mathrm{PNS}$ are those at  $t_\mathrm{pb}=t_\mathrm{pb,\;400}$. The columns for the ALP cooling power $L_a$ and the heating power $L_\mathrm{heat}$ show their values at $t_\mathrm{pb}=200$ ms.  The row with $g_{10}=0$ represents the 
model without ALPs.}
\end{table*}

We perform {two core-collapse simulations} without ALPs  and {24} SN simulations with $m_a=50$--$800$\,MeV and $g_{10}=g_{a\gamma}/(10^{-10}\;\mathrm{GeV}^{-1})=4$--$40$. Our focus on this range is motivated by studies employing the post-processing method to estimate the ALP effects and find the energy deposited by ALPs behind the shock can be 
very high 
\cite{2020JCAP...12..008L}. {The SN models developed in this study are listed in Table I.}
The parameter region explored by these models has not been excluded by previous works on the SN 1987A limits \citep{2018PhRvD..98e5032J,2020JCAP...12..008L}. 

  \begin{figure}
  \centering
  \includegraphics[width=8cm]{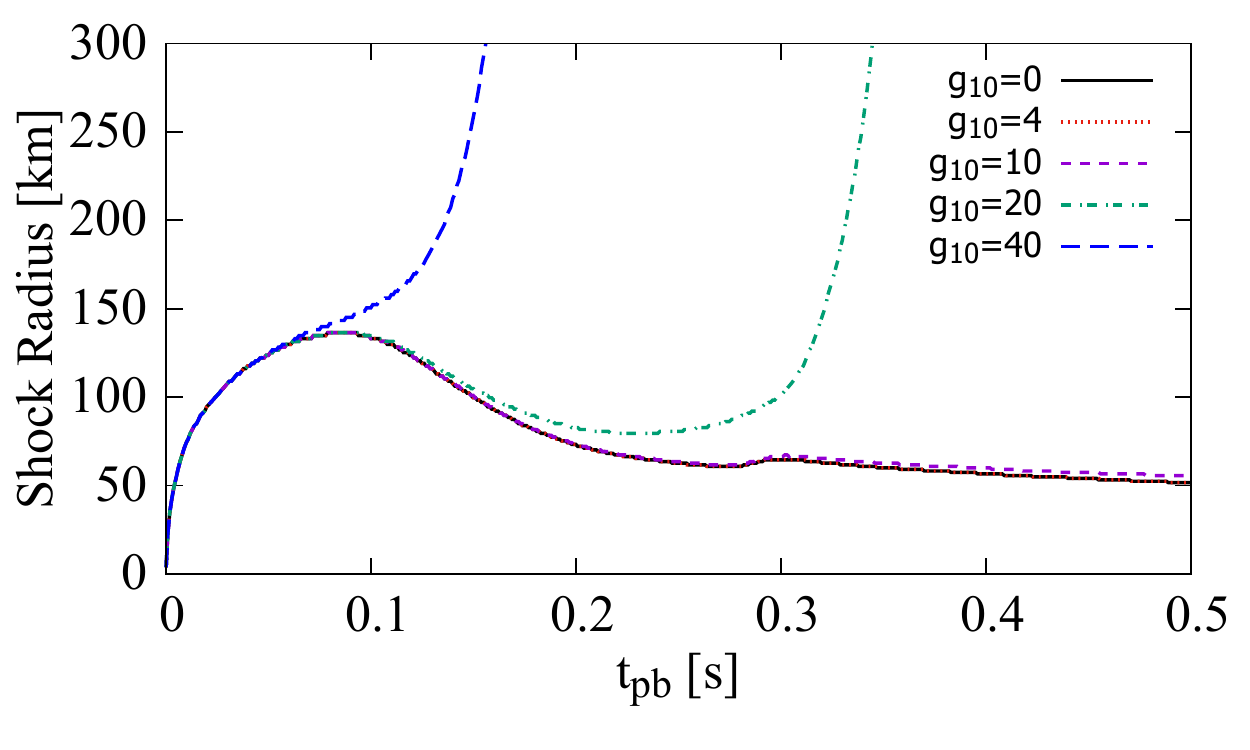}
  \caption{The radius of the shock wave {in the $20M_\odot$ model} as functions of time $t_\mathrm{pb}$ since the core bounce. Shown are models including ALPs with $m_a=100$\,MeV and different couplings, as labeled, as well as the model without ALPs (black solid) for comparison.}
  \label{shock}
 \end{figure}
 
  \begin{figure}
  \centering
  \includegraphics[width=8cm]{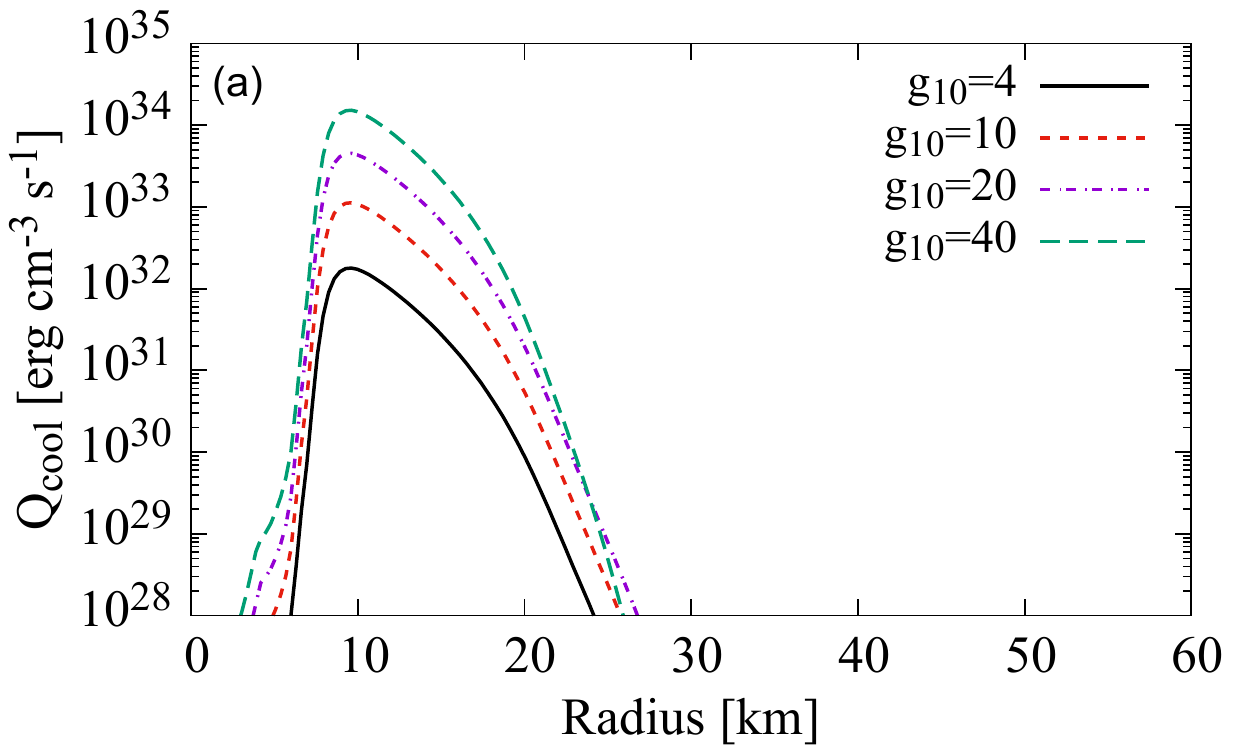}
   \includegraphics[width=8cm]{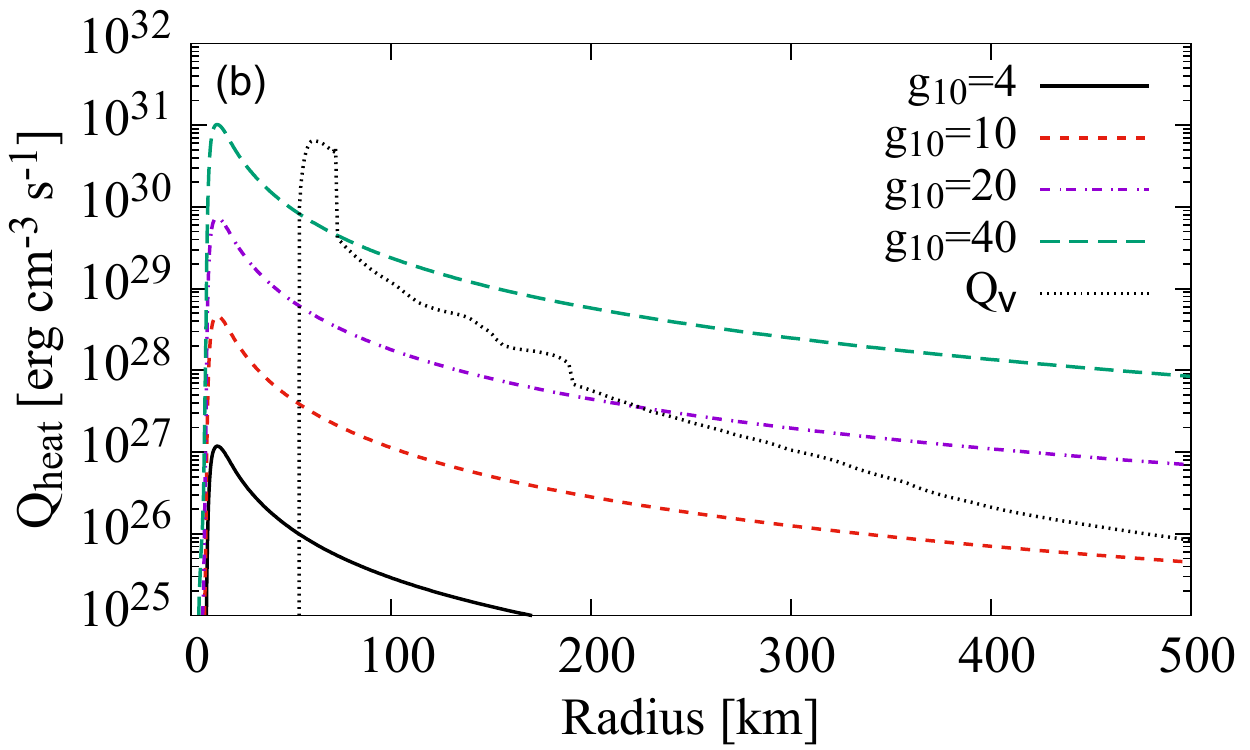}
  \caption{The radial profiles of (a) ALP cooling rate $Q_\mathrm{cool}$,  and (b) ALP heating rate $Q_\mathrm{heat}$, both {in the $20M_\odot$ model} at $t_\mathrm{pb}=200$ ms. The ALP mass is fixed to $m_a=100$\,MeV. In the bottom panel, the black dotted line shows the net neutrino energy $Q_\nu=Q^\nu_\mathrm{heat}-Q^\nu_\mathrm{cool}$ in the $g_{10}=0$ model.}
  \label{Q}
 \end{figure}

  In the neutrino-driven explosion scenario, neutrinos heat up the matter behind the shock wave and help the explosion. Since most one-dimensional SN models fail to explode, it is usually argued that multi-dimensional effects including convection and the standing accretion shock instability are essential to achieve successful explosions \citep[e.g.][]{2007ApJ...656.1019Y,2013ApJ...765..110D,2013ApJ...775...35C,2013ApJ...770...66H,2018ApJ...865...81O}. This argument is supported by recent multi-dimensional simulations, some of which reproduce $10^{51}$ erg explosions \cite[e.g.][]{2017MNRAS.472..491M,2018ApJ...855L...3O,2020MNRAS.491.2715B,2021ApJ...915...28B}. However,  ALPs can heat the matter as well and potentially lead to shock revival even in one-dimensional models.
  
  \subsection{$20M_\odot$ Models}
  {In this study, we adopt the $20M_\odot$ models as a fiducial case.} Figure~\ref{shock} shows the time evolution of the shock wave  after the core bounce in models with $m_a=100$\,MeV. We employed coupling constants $g_{10}=0$, 4, 10, 20, and 40, where the model with $g_{10}=0$ is the standard model without ALPs. In the cases of $g_{10}=0$, 4, and 10, the shock wave 
  never exceeds 
  $r\sim 150$\,km. 
  This stalling is similar to conventional one-dimensional models. However, the shock wave is revived in the cases of $g_{10}=20$ and 40 because $Q_\mathrm{heat}$ due to ALPs are large enough in these models. This provides a new scenario to reproduce successful SN explosions even in spherically-symmetric cases.

 In Fig.~\ref{Q}, we show the ALP cooling rate $Q_\mathrm{cool}$ and the heating rate $Q_\mathrm{heat}$ as a function of radius. The time since the core bounce is fixed to $t_\mathrm{pb}=200$\,ms and the ALP mass is fixed to $m_a=100$\,MeV. It is seen that the ALP production is localised at $r\sim 10$\,km. This is because the temperature is the highest in this region and the ALP production rate is a steep function of $T$. The values of $Q_\mathrm{cool}$ increase as a function of $g_{10}$ because the ALP production rates shown in Eqs.~(\ref{prim}) and (\ref{pc}) are proportional to $g_{10}^2$. In the models with $m_a\geq 100$\,MeV, the contribution of the photon coalescence is larger than that of the Primakoff process, while the contribution of the Primakoff process is dominant when $m_a=50$ MeV.

 Once produced at $r\sim10$\,km, ALPs propagate through the SN matter. A part of the ALPs is absorbed and heats up the fluid. Figure~\ref{Q}(b) shows the ALP heating rate {for the $m_a=100$ MeV models} at $t_\mathrm{pb}=200$\,ms. The heating rate is the largest at $r\sim 15$\,km and decreases in outer regions. The values of $Q_\mathrm{heat}$ is approximately proportional to $g_{10}^4$ because the number of produced ALPs is proportional to $g_{10}^2$ and the radiative decay rate  is also proportional to  $g_{10}^2$ as we see from  Eq.~(\ref{decay}). {Table I shows the ALP heating power 
 \begin{eqnarray}
 L_\mathrm{heat}=4\pi\int_{0\;\mathrm{km}}^{5000\;\mathrm{km}} Q_\mathrm{heat}r^2dr
 \end{eqnarray} 
 at $t_\mathrm{pb}=200$\,ms. The integration is performed over the range of the simulation, i.e. $r\in[0\;\mathrm{km},\;5000\;\mathrm{km}]$. Since the simulation is only for the SN core, $L_\mathrm{heat}$ in Table I does not include the deposited energy in the stellar envelope.} When $m_a=50$ MeV, $L_\mathrm{heat}$ is approximately proportional to $g_{10}^4$. The dependence of $L_\mathrm{heat}$ on $g_{10}$ becomes weaker for heavier ALPs. When ALPs are as heavy as 800 MeV, one can see that $L_\mathrm{heat}\propto g_{10}^2$. This is because the MFP is so short that ALPs are completely trapped by the star and hence $L_\mathrm{heat}\approx L_a$, where $L_a$ is the ALP luminosity defined in Eq. (\ref{Lcool}).
  \begin{figure}
  \centering
  \includegraphics[width=8cm]{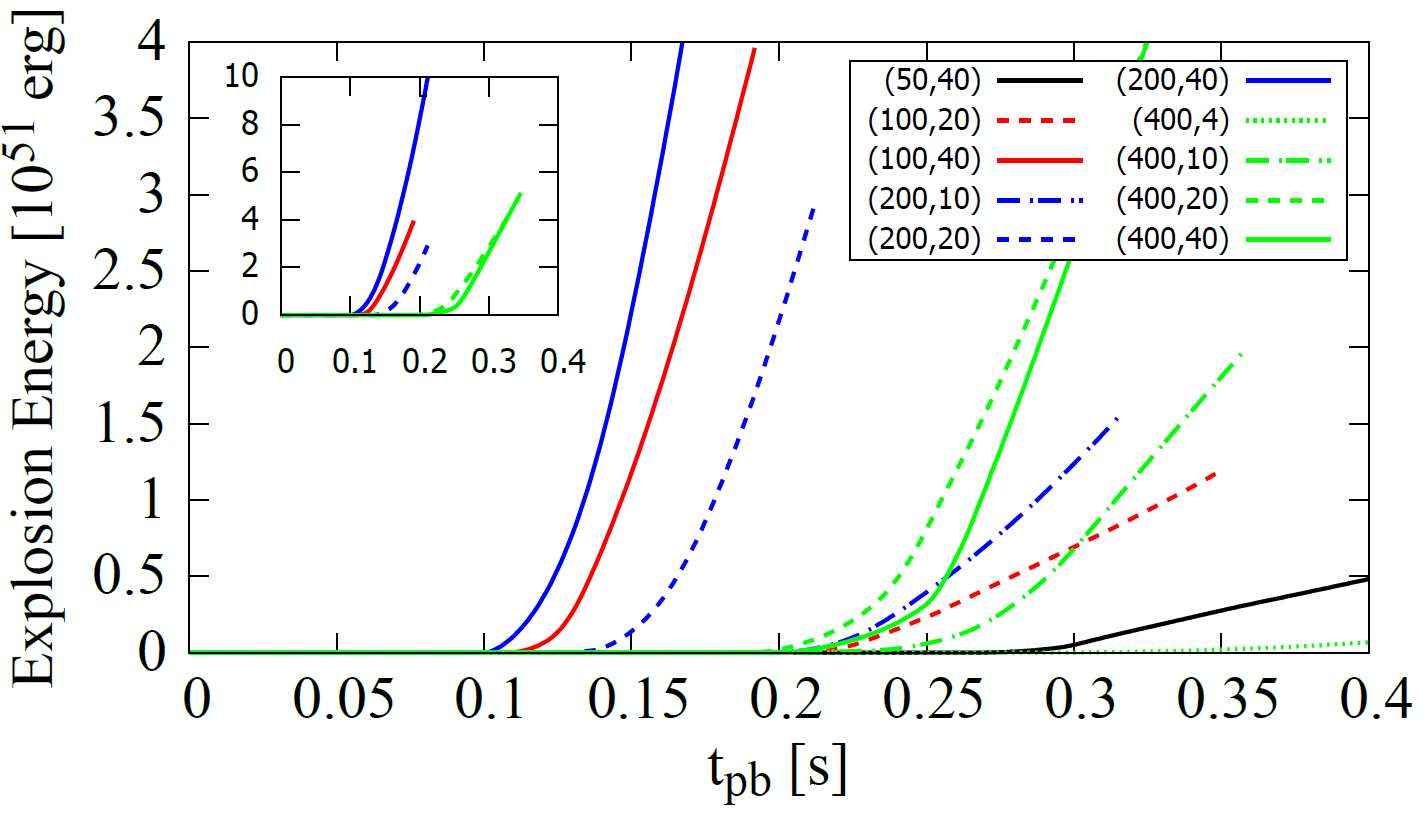}
  \caption{The explosion energies of all {$20M_\odot$} models that achieved successful explosions as a function of the time $t_\mathrm{pb}$ since the core bounce. Each model is designated by a pair of $(m_a/1\;\mathrm{MeV},\;g_{10})$. The inset panel shows the explosion energies of energetic models that exceed $E_\mathrm{exp}>4\times10^{51}$ erg.}
  \label{Eexp}
  \end{figure}
  
  Fig.~\ref{Eexp} shows the explosion energy \cite{2011ApJ...738..165S}
  \begin{eqnarray}
  E_\mathrm{exp}=\int_D dV\left(\frac{1}{2}\rho v^2+e-\rho\Phi\right)
  \end{eqnarray}
  for the models with successful explosions. Here $v$ is the velocity, $e$ is the internal energy, and $\Phi$ is the gravitational potential. The region $D$ is the domain where the integrand is positive. In the figure, each model is designated by a pair of $(m_a/1\;\mathrm{MeV},\;g_{10})$. In all models except for the ones with $m_a=400$\,MeV, the explosions start earlier when $g_{10}$ is larger because of higher values of $Q_\mathrm{heat}$. In the model of $m_a=400$\,MeV and $g_{10}=40$, the MFP of ALPs is as short as $\sim 23$\,km. In this case, the MFP is so short that the shock wave is not heated effectively and consequently the monotonous dependence on $g_{10}$ is not observed for $m_a=400$\,MeV.
    \begin{figure}
  \centering
  \includegraphics[width=8cm]{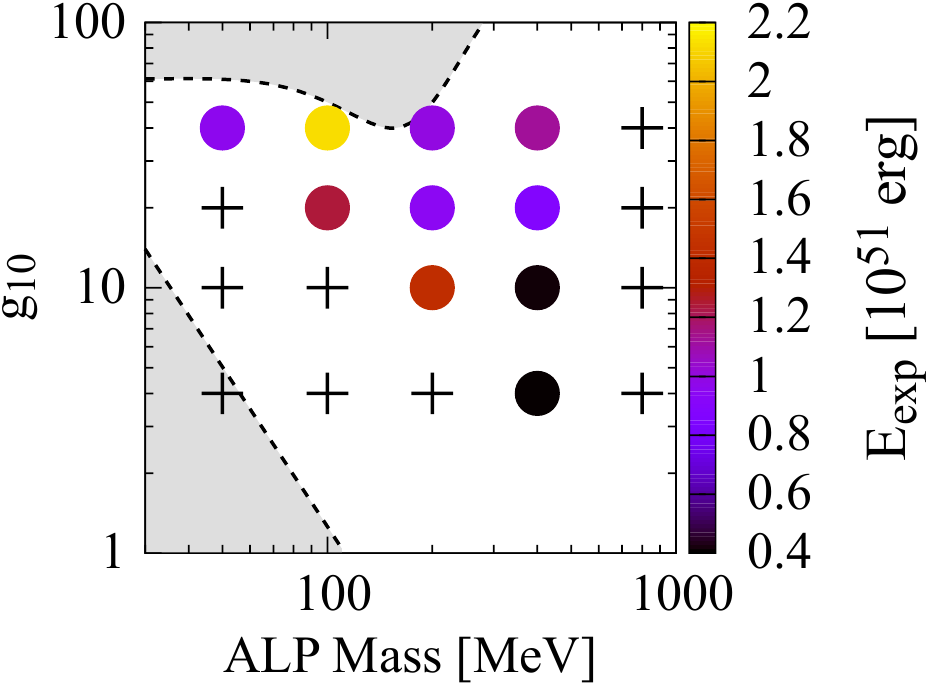}
  \caption{Outcome of core collapse {of the $20M_\odot$ star} for various combinations of ALP parameters adopted in this work. The crosses represent models which failed to explode while filled circles represent models which successfully explode. The color shows the explosion energy in units of $10^{51}$ erg at the moment when the shock wave reaches $r=400$ km. The grey regions show the SN 1987A limits \cite{2018PhRvD..98e5032J,2020JCAP...12..008L}.}
  \label{Eexp_param}
 \end{figure}
 
  The simulations are stopped before $E_\mathrm{exp}$ saturates. Even so, $E_\mathrm{exp}$ of the model with $(m_a/1\;\mathrm{MeV},\;g_{10})=(200,\;40)$ already exceeds $10^{52}$ erg. Also, $E_\mathrm{exp}$ of the models  with $(m_a/1\;\mathrm{MeV},\;g_{10})=(100,\;40)$, (200, 20), (400, 40), and (400, 20) would eventually exceed $10^{52}$\,erg. Given these high explosion energies, these models might be observed as broad-line type Ic SNe, whose mechanism is still under debate. Other models with $E_\mathrm{exp}\sim10^{51}$\,erg would be candidates for ordinary SN explosions. 
 
 Figure~\ref{Eexp_param} shows $E_\mathrm{exp}$ at the moment when the shock wave reaches $r=400$ km. Here, crosses represent models with failed explosion, i.e., the SN shock is not energetically revived. The explosion energies tend to increase as a function of $g_{10}$ because higher values of $Q_\mathrm{heat}$ are obtained. In general, heavier ALPs are conducive to  explosions. This is intuitively explained by the fact that the MFP of heavier ALPs is shorter and hence the shock wave is heated more efficiently. This trend breaks down at $m_a = 800$\,MeV, because the temperature in the proto-neutron star is not high enough to produce such heavy ALPs.

{The ALP heating affects the proto-neutron star mass $M_\mathrm{PNS}$ as well. In the standard one-dimensional model, the shock revival does not occur and the mass accretion on the proto-neutron star does not stop. As a result, the star implodes to leave a black hole. On the other hand, Table I shows the values of $M_\mathrm{PNS}$ in each model with successful explosion at the moment when the shock wave reaches $r=400$\,km. The remnant mass becomes smaller with the effect of ALPs because additional heating prevents the mass accretion. In our parameter region, the models show $M_\mathrm{PNS}\sim(1.6$--$1.9)M_\odot$, which is below the maximum mass of neutron stars. Hence these models will leave a neutron star as a remnant. }
     \begin{figure}
  \centering
  \includegraphics[width=8cm]{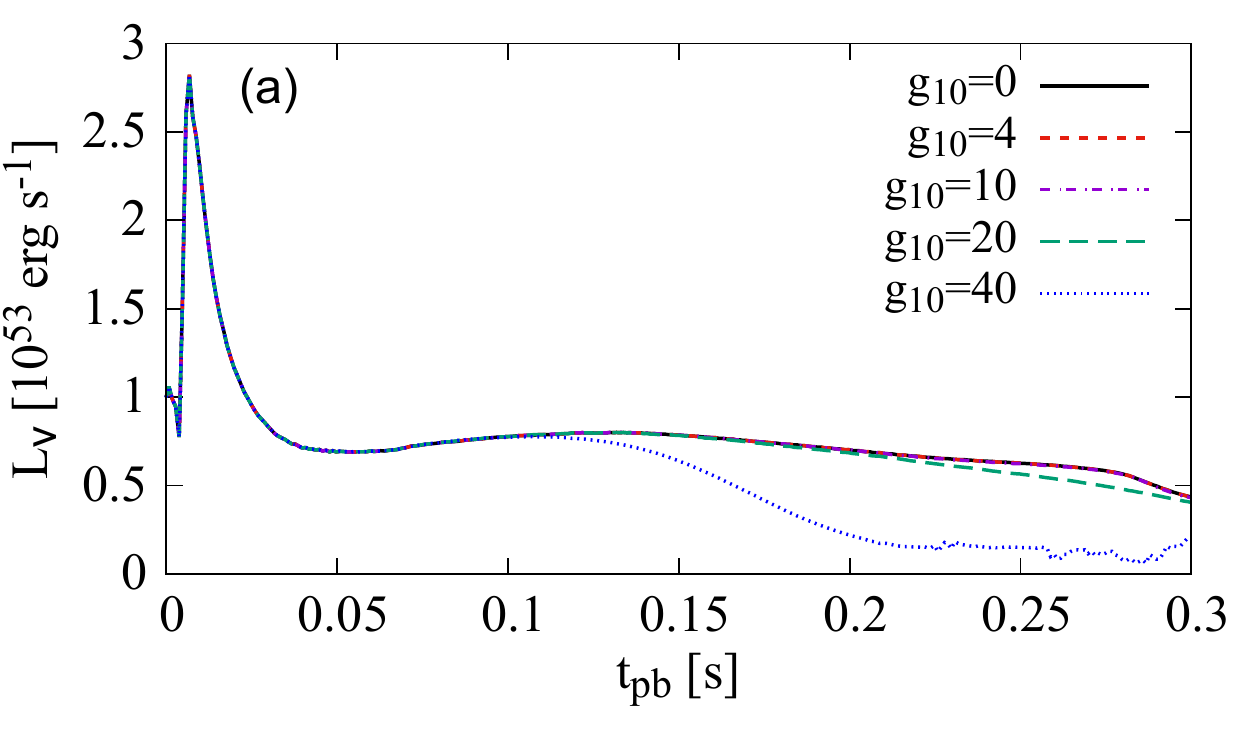}
    \includegraphics[width=8cm]{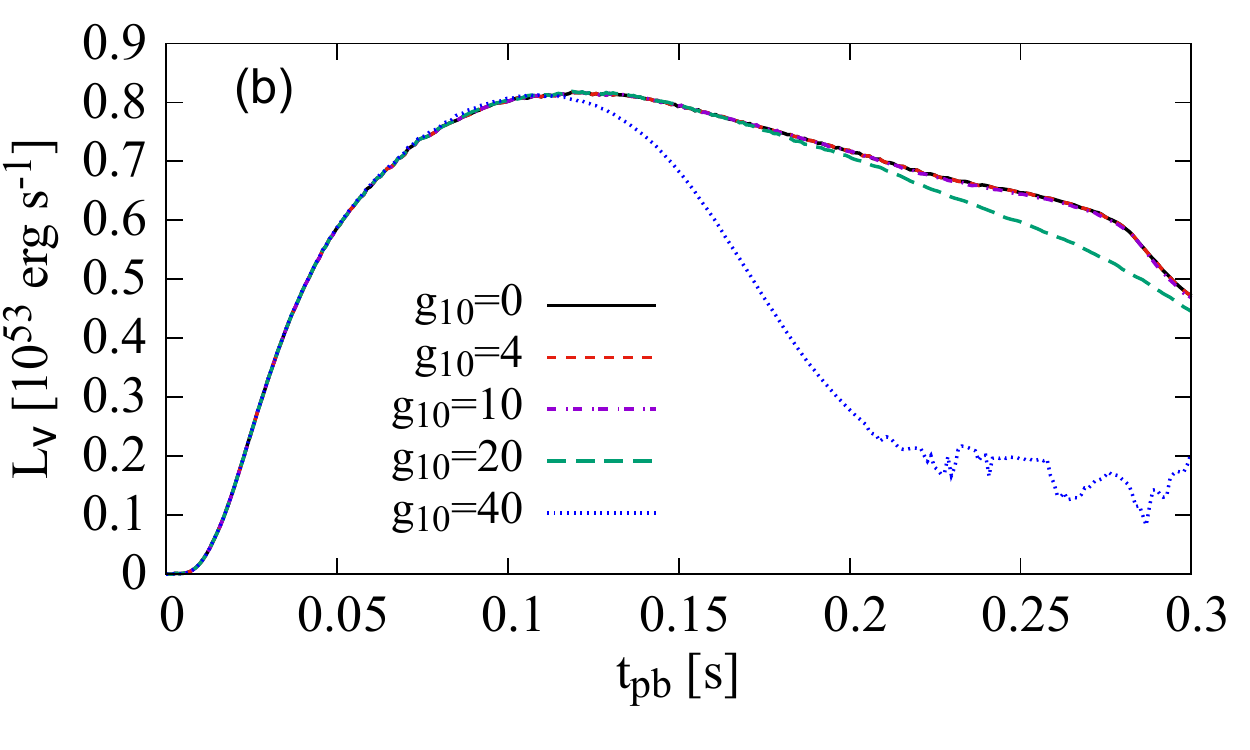}
  \caption{The neutrino luminosities in the {$20M_\odot$} models with $m_a=100$ MeV. The panel (a) is for electron neutrinos and (b) is for electron anti-neutrinos.}
  \label{Ln}
 \end{figure}
 
 {Fig. \ref{Ln} shows the $\nu_e$ and $\bar{\nu}_e$ luminosities $L_\nu$ for the models with $m_a=100$ MeV. Electron neutrinos are mainly produced by the electron capture of protons. As a consequence,  the neutronization burst is observed when the shock wave reaches the neutrino sphere. On the other hand, the $\bar{\nu}_e$ luminosity starts increasing after the $\nu_e$ luminosity because electron anti-neutrinos are produced by other processes such as the positron capture of neutrons, nucleon bremsstrahlung, and electron-positron annihilation.  If $g_{10}\lesssim10$, the neutrino luminosities are not affected by ALPs, while the luminosities are smaller than the standard model without ALPs  if $g_{10}$ is larger. This is because the accretion on the proto-neutron star stops in the models with successful explosions.}

\subsection{Dependence on the Progenitor Models}
{As shown in Table~I, we developed 5 models with the $11.2M_\odot$ progenitor for $m_a=100$ MeV.  It is found that ALP heating can assist the shock revival as well when $g_{10}$ is sufficiently high. Compared with the corresponding $20M_\odot$ models, the explosion energy in the $11.2M_\odot$ models is lower. This is because the peak temperature in the proto-neutron star is 51.4\,MeV in the $20M_\odot$ models, while its value in the $11.2M_\odot$ models is as low as 35.4\,MeV. As a result, the heating power $L_\mathrm{heat}$ in the lighter models are $\sim 3$--$5$ times lower than those in the heavier models. Also, the mass of the proto-neutron star becomes smaller because of ALPs. It is hence expected that the mass distribution of neutron stars can be affected by ALPs, although it is beyond the scope of this study to perform thorough investigation on stellar mass dependence.}

 \section{Discussion and Conclusions}
In this study, we performed consistent SN simulations coupled with the production and absorption of heavy ALPs. {We showed that} if the mass of ALPs is $50$--$400$ MeV and the ALP-photon coupling constant $g_{10}$ is sufficiently high, the shock wave is efficiently heated and successful explosions are obtained. {In a recent study \cite{2020JCAP...12..008L}, it was pointed out that the ALP luminosity from the neutrino sphere can be powerful, but using the post-processing technique. Our study serves as a numerical confirmation of this and supports the potential triggering of explosions.}

{Some of our ALP parameters} resulted in energetic explosions reaching $E_\mathrm{exp}\sim10^{52}$\,erg. Such models {are reaching the explosion energetics of} broad-line type Ic SNe. 
{Assuming the universality of ALP parameters, this would predict that} most of SNe would be as energetic as $\sim10^{52}$\,erg. 
{Since it} is observationally estimated that broad-line type Ic SNe represent only $\sim 2\%$ of all CCSNe in giant galaxies and $\sim 13\%$ in dwarf galaxies \citep{2010ApJ...721..777A}, {such ALP parameter regions should} 
be regarded as excluded on the basis of the explosion energy of the typical SNe \cite{2019PhRvD..99l1305S}. 
{On the other hand, the models with ($m_a$, $g_{10}$)=(400 MeV, 4) and (50 MeV, 40) interestingly show a moderate explosion which is similar to the majority of SNe.}
{Future} work are needed to understand the population statistics, including dependence on progenitors and nuclear equations of state, to clarify observational consequences of ALP heating and setting limits on ALP parameters.

After this paper was submitted, Ref.~\cite{2022arXiv220109890C} appeared in arXiv. This paper
argues that the ALP parameters which lead to $E_\mathrm{exp}>10^{50}$\,erg should be excluded on the basis of observed low-energy SNe, although their calculation adopted post-processing. If we adopt their criterion, all of our models with successful explosion are excluded because they show $E_\mathrm{exp}>10^{50}$\,erg. The excluded region for $m_a\sim400$\,MeV based on our simulations seems to be slightly broader than the result in Ref.~\cite{2022arXiv220109890C}. This difference may be attributed to the difference in progenitors and microphysics such as equations of state.  Also, Ref.~\cite{2022arXiv220109890C} implies that our models  with $m_a\leq200$\,MeV that fail to revive the shock wave are excluded as well. This argument cannot be confirmed by our simulations because the simulated region is $r<5000$\,km.

{There are many potential consequences of the ALP explosion scenario, some of which we briefly discuss here. For example,} ALPs may affect SN r-process nucleosynthesis. Although CCSNe used to be a candidate of r-process site \cite[e.g.][]{2004A&A...416..997A}, the r-process is suppressed in recent standard models \citep{2010A&A...517A..80F,2013ApJ...770L..22W} because the irradiation by neutrinos creates a neutron poor composition. Nevertheless, if the additional ALP heating causes an earlier shock revival, the neutrino irradiation may be reduced and the outflow may maintain a neutron rich composition, providing helpful conditions for r-process nucleosynthesis. Recently, the effect of the hadron-quark phase transition on SN explosion has been investigated \cite{2018NatAs...2..980F,2020PhRvL.125e1102Z,2021arXiv210901508K}. Contrary to the standard scenario, r-process can occur in such models \cite{2020ApJ...894....9F}.  Since the dynamics of our model with ALPs is similar to these models, SN explosions induced by ALP heating might work as an r-process site.
{As another consequence,} ALPs may affect the remnant of the stellar core collapse. Because ALPs can help shock revival, accretion on a proto-neutron star may be suppressed, compared with the standard models without ALPs. As a result, a neutron star can remain in an SN remnant instead of a black hole. This effect might affect the mass functions of neutron stars and black holes. It is desirable to perform systematic calculations with ALPs for a wide range of progenitors to understand the effect on remnants.

{The ALP emission also influences the neutrino emission, and extreme effects can rule out ALP parameter space.} Ref.~\cite{2020JCAP...12..008L} argued that the ALP luminosity at $t_\mathrm{pb}=1$\,s should not exceed the neutrino luminosity $\sim 3\times10^{52}$ erg s$^{-1}$ from SN 1987A and constrained the ALP parameters. Since our calculations focus on explosion dynamics at $t_\mathrm{pb}\lesssim0.5$\,s, our results do not contradict their argument. However, it is desirable to perform long-time simulations coupled with ALPs to verify their constraints because they have neglected the back-reaction of additional cooling and heating.
{Similarly, ALPs may affect gravitational wave signals as well. Hence, predictions of multi-messenger signals could provide unique connections with observables of future nearby SNe. While beyond the scope of this study, multi-dimensional simulations coupled with ALPs are therefore indispensable to predict detailed neutrino and gravitational wave signals.}

Besides astrophysical arguments, cosmology also provides probes of ALPs \cite[e.g.][]{1992SvJNP..55.1063B,2012JCAP...02..032C,2020JCAP...05..009D}. ALPs could be produced in the Early Universe when electrons and neutrinos are coupled. If ALPs decay after decoupling with electrons and neutrinos, neutrinos would have smaller energies than in the standard cosmological model and the number of effective neutrinos. Also, ALPs would inject photons which may dilute the neutrino and baryon densities in the epoch of Big Bang nucleosynthesis and affect primordial elemental abundances. {Although a part of the ALP parameter space focused in this study is excluded by cosmological arguments, some of the parameter space remains. Also, as described in Ref.~\cite{2020JCAP...12..008L}, cosmological constraints come with additional assumptions about the cosmological model. For example, a conservative assumption about the reheating temperature may render cosmological constraints weaker by orders of magnitude \cite{2020JCAP...12..008L}. Thus, it is worthwhile to explore the parameter space with independent methods to exclude the effects of systematic uncertainties.}

In our simulations, we focused on ALPs that couple only with photons. However, ALPs can couple with other standard model particles such as nucleons \cite[e.g.,][]{PhysRevLett.60.1793,PhysRevD.39.1020,2016PhRvD..94h5012F,2019JCAP...10..016C,2021PhRvD.104j3012F} and electrons \cite{2021PhRvD.104j3007L,2021PhRvD.104d3016C}. {It has been reported that CCSNe provide information on these
couplings as well,
although their backreaction on hydrodynamics in the context of multi-dimensional
simulations
has not been explored in detail.} 
It would be desirable to develop SN models to explore how additional couplings impact the energy transport in SNe. For example, since ALP-nucleon coupling can produce significantly more ALP luminosity than ALP-photon coupling \cite[e.g.,][]{2020PhRvD.102l3005C}, this combination could be more potent than ALP-photon coupling alone.

{In summary, we explored numerically the generation and absorption of heavy ALPs self-consistently in the collapse of massive stars. We showed that ALPs can be produced in large quantities and can trigger powerful explosions. However, more studies, covering both time and progenitor diversity, are needed to fully explore the viability of the heavy ALP scenario. Nevertheless, our study supports the intriguing possibility of ALP triggered explosions. Future systematic studies may make connections with SN observables while at the same time allow to place constraints on ALPs.} 

\begin{acknowledgments}
This work is supported by Research Institute of Stellar Explosive Phenomena at Fukuoka University and JSPS KAKENHI Grant Numbers JP21K20369, JP17H06364, JP18H01212, and JP21H01088. Numerical computations were  carried out on the PC cluster at Center for Computational Astrophysics, National Astronomical Observatory of Japan.
The work of S.H.~is supported by the U.S.~Department of Energy under the award number DE-SC0020262 and NSF Grant numbers AST-1908960 and PHY-1914409. This work was supported by World Premier International Research Center Initiative (WPI Initiative), MEXT, Japan. 
\end{acknowledgments}

\appendix*
\section{Comparison of methods to calculate the heating rate}
 \begin{figure}[t]
 \includegraphics[width=8.5cm]{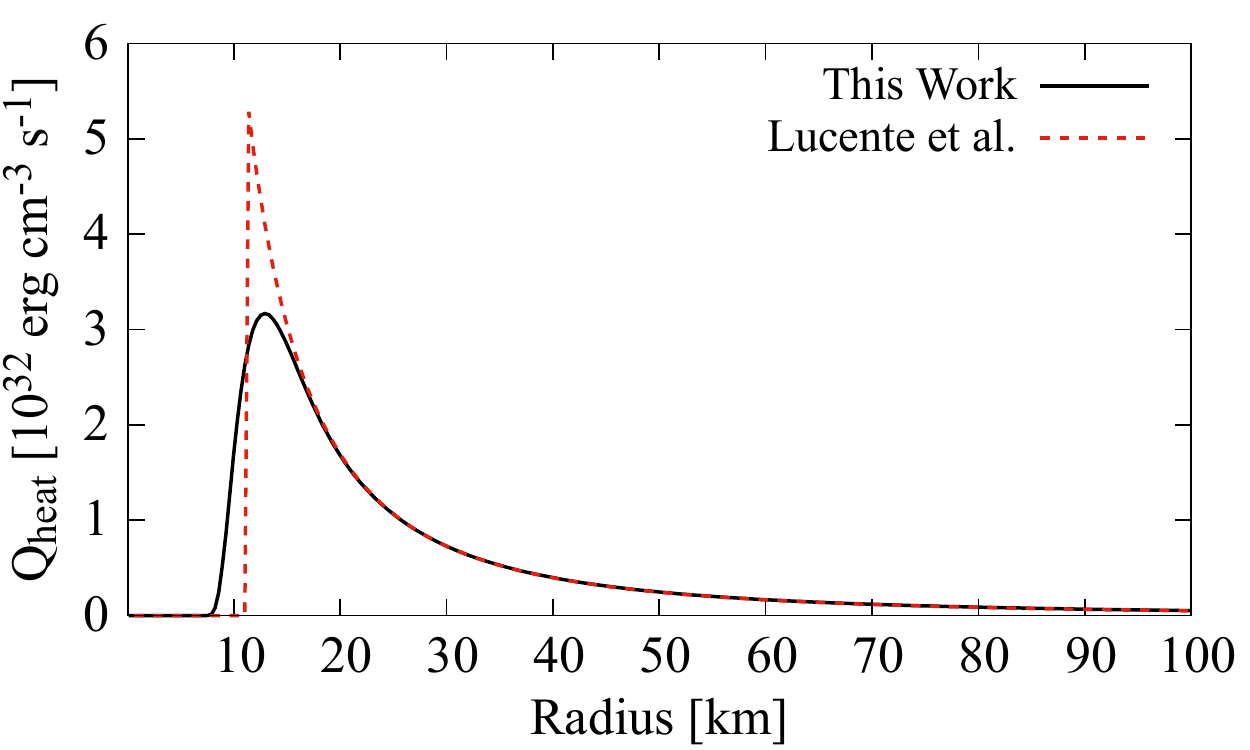}
 \caption{The ALP heating rate $Q_\mathrm{heat}$ at $t_\mathrm{pb}=200$ ms as a function of the radius. The solid line adopts Eq. (\ref{qheat}) and the broken line adopts Eq. (\ref{qheat2}). The ALP mass is fixed to $m_a=200$ MeV and the ALP-photon coupling is fixed to $g_{10}=40$. The SN model does not include the effect of ALPs and the calculation of $Q_\mathrm{heat}$ is performed as post-processing.}
 \label{comparison}
\end{figure}
{In this study, we calculated the ALP heating rate $Q_\mathrm{heat}$ on the basis of a recurrence relation (Eq. (\ref{rec})) between successive  cells. However, Ref. \cite{2020JCAP...12..008L} adopted another method to calculate $Q_\mathrm{heat}$. In this appendix, we compare results of the two methods. }

{Ref. \cite{2020JCAP...12..008L} defined the ALP optical depth $\tau_a(r,\;R)$ between radii $R$ and $r$ as
\begin{eqnarray}
\tau_a(r,\;R)=\int^{R}_r\frac{d\tilde{r}}{\lambda_a(\langle E_a\rangle,\;\tilde{r})},\label{tau}
\end{eqnarray}
where $\lambda_a(\langle E_a\rangle,\;\tilde{r})$ is the MFP of ALPs at $\tilde{r}$ with the averaged ALP energy $\langle E_a\rangle$. Using $\tau_a$, the  energy deposited by ALPs at $R$ per unit time is written as
\begin{eqnarray}
L_\mathrm{dep}(t,\;R)=L_a(t)(1-\exp(-\tau_a(R_p,\;R))).\label{Ldep}
\end{eqnarray}
Here $L_a(t)$ is the ALP luminosity defined as
\begin{eqnarray}
L_a(t)=4\pi\int Q_\mathrm{cool}r^2dr \label{Lcool}
\end{eqnarray}
and $R_p$ is the mean radius of the ALP production that is calculated as
\begin{eqnarray}
R_p=\frac{\int rQ_\mathrm{cool}dr}{\int  Q_\mathrm{cool}dr}.
\end{eqnarray}
Since 
\begin{eqnarray}
L_\mathrm{dep}(t,\;R)\approx 4\pi R^2(\Delta R)Q_\mathrm{heat},\label{qheat2}
\end{eqnarray}
we can calculate the ALP heating rate $Q_\mathrm{heat}$ from the energy deposition rate defined in eq. (\ref{Ldep}). }

{Figure \ref{comparison} shows the comparison of the ALP heating rates $Q_\mathrm{heat}$ calculated with Eqs. (\ref{qheat}) and (\ref{qheat2}).  Here the ALP mass is fixed to $m_a=200$ MeV and the ALP-photon coupling is fixed to $g_{10}=40$. In this calculation, we adopt the standard SN model without the effects of ALPs and $Q_\mathrm{heat}$ is treated as a post-process. When $Q_\mathrm{heat}$ is calculated with the previous method, $Q_\mathrm{heat}$ increases discontinuously at $R_\mathrm{p}\approx11$ km, while it rises smoothly in our method. Although the two methods are apparently different, the results for $Q_\mathrm{heat}$ coincide with each other at $r\gtrsim20$ km. }
\bibliography{ref.bib}
\end{document}